\documentclass[twocolumn,showpacs,amsmath]{revtex4-1}
\usepackage{amsmath,amssymb,amsfonts,bbm,graphicx}
\usepackage[latin1]{inputenc}
\usepackage[T1]{fontenc}
\usepackage{verbatim}
\usepackage{textcomp}
\usepackage{graphicx}
\usepackage{amsmath}

\usepackage{color}

\begin{document}
\title{First-principles study of the ferroelectric Aurivillius phase Bi$_2$WO$_6$}
\author{Hania Djani$^{1,2}$, Eric Bousquet$^{3,4}$, Abdelhafid Kellou$^2$, Philippe Ghosez$^3$}
\affiliation{
$^1$Centre de D\'eveloppement des Technologies Avanc\'ees, cit\'e 20 aout 1956, Baba Hassen, Alger, Algeria \\
$^2$Facult\'e de Physique, USTHB, BP.32 El Alia, Bab Ezzouar, 16111 Alger, Algeria\\
$^3$Physique Th\'eorique des Mat\'eriaux, Universit\'e de Li\`ege (B5), B-4000 Li\`ege, Belgium\\
$^4$Department of Materials, ETH Zurich, Wolfgang-Pauli-Strasse 27, CH-8093 Zurich, Switzerland\\}


\begin{abstract}
In order to better understand the reconstructive ferroelectric-paraelectric transition of Bi$_2$WO$_6$, which is unusual within the Aurivillius family of compounds, we performed first principles-calculations of the dielectric and dynamical properties on two possible high-temperature paraelectic structures: the monoclinic phase of $A2/m$ symmetry observed experimentally and the tetragonal phase of $I4/mmm$ symmetry, common to most Aurivillius phase components. Both paraelectric structures exhibit various unstable modes, which after their condensation bring the system toward more stable structures of lower symmetry. The calculations confirm that, starting from the paraelectric $A2/m$ phase at high temperature, the system must undergo a reconstructive transition to reach the $P2_1ab$ ferroelectric ground state.
\end{abstract}
\maketitle

\section{Introduction}

The Aurivillius phases constitute a family of naturally occurring complex layered oxide structures in which  fluorite-like (Bi$_2$O$_2$)$^{+2}$ units alternate with perovskite-like (A$_{m-1}$B$_{m}$O$_{3m+1}$)$^{-2}$ blocks, where A= Ca$^{+2}$, Ba$^{+2}$, Sr$^{+2}$, Bi$^{+2}$, etc., B = Fe$^{+3}$, Ti$^{+4}$, Nb$^{+5}$, Ta$^{+5}$, W$^{+6}$ , etc. and $m$ is the number of BO$_6$ octahedra in the perovskite-like blocks ($m=1 - 8$) \cite{1}. The oxidation states of the cations and $m$ are interdependent to always provide the right number of electrons to oxygen atoms. This family has numerous representatives, the most studied being Bi$_2$WO$_6$ ($m=1$), SrBi$_2$Ta$_2$O$_9$ ($m=2$) and Bi$_4$Ti$_3$O$_{12}$ ($m=3$) \cite{2, 3, 4, 5, 6}.

  The ferroelectric properties of the Aurivillius phases have been widely studied since the early works of Subbarao \cite{7}  and  Smolenski \textit{et al.} \cite{8} and they continue to rise technological interest mainly for their potential use in thin  film non-volatile memory applications \cite{9}. They are also attractive  at the fundamental level because of the peculiarity of their phase transitions and the complex interplay of their  structural instabilities.

 Bi$_2$WO$_6$ is the simplest member of the Aurivillius family with one oxygen octahedron sandwiched between two Bi$_2$O$_2$ layers. It possesses many interesting physical properties such as ferroelectricity with large spontaneous polarization and  high Curie temperature (950\textdegree C), piezoelectricity with a potential for high-frequency and high-temperature applications  as an alternative to BaTiO$_3$ and PbZr$_{1-x}$Ti$_{x}$O$_3$ \cite{10}, high ion conductivity \cite{11} and photocatalytic activity \cite{12}.

Understanding ferroelectricity and the sequence of phase transitions in this compound is still challenging. Experimental studies \cite{13}  showed that Bi$_2$WO$_6$ is polar at ambient temperature with orthorhombic $P2_1ab$ (No. 29) structure. Rising the temperature reveals at 670\textdegree C a transition to an intermediate polar phase with orthorhombic $B2cb$ (No. 41) structure. This structure is maintained until 950\textdegree C, then the material displays a transition to a paraelectric monoclinic $A2/m$ (No. 12) structure.
This type of ferroelectric-paraelectric transition to an $A2/m$ phase is unexpected and rather unique among Aurivillius phases, where usually compounds transform at high temperature into a centrosymmetric tetragonal $I4/mmm$ (No. 139) structure \cite{6}. The same kind of behavior appears in Bi$_2$MoO$_6$, a compound isomorphic to Bi$_2$WO$_6$, which  exhibits orthorhombic structures at low and intermediate temperatures and a monoclinic $P2_1/c$ (No. 14) structure at high temperature \cite{14,15}. Such ferroelectric to paraelectric transitions from orthorhombic to monoclinic are necessarily reconstructive since the space groups of the low temperature phases are not subgroups of the high temperature phases ($A2/m$ or $P2_1/c$), but rather of the $I4/mmm$ structure.

We performed first-principles calculations in order to better understand these intriguing observations. First-principles studies of Bi$_2$WO$_6$ have been previously reported in the literature. Machado {\it et al.} \cite{16} performed frozen-phonon calculations of the polar modes at the Brillouin-zone center in the hypothetical I4/mmm tetragonal paraelectric phase in order to clarify the microscopic origin of the ferroelectric instability. More recently, Mohn and Stolen \cite{17} highlighted the modification of the oxygen arrangement around Bi atoms in the Bi$_2$O$_2$ layers at the reconstructive ferroelectric-paraelectric transition. In the present work, we complete these studies by reporting a systematic investigation of the structural, dielectric and dynamical properties of both potential high-temperature A2/m and I4/mmm paraelectric phases. We investigate the energy landscape around these two phases. The structure and internal energy of various other intermediate phases, reached by the condensation of individual or combined lattice instabilities at the zone center and zone boundaries of the Brillouin zone, are calculated. A diagram comparing the internal energy of various possible phases is then established, which confirms that starting from high-temperature paraelectric $A2/m$ phase, the system must undergo on cooling a reconstructive phase transition to reach the $P2_1ab$ ground-state structure. The spontaneous polarizations of different polar phases are also reported.    
\begin{figure}[t]
\centering\includegraphics[angle=0,scale=0.090]{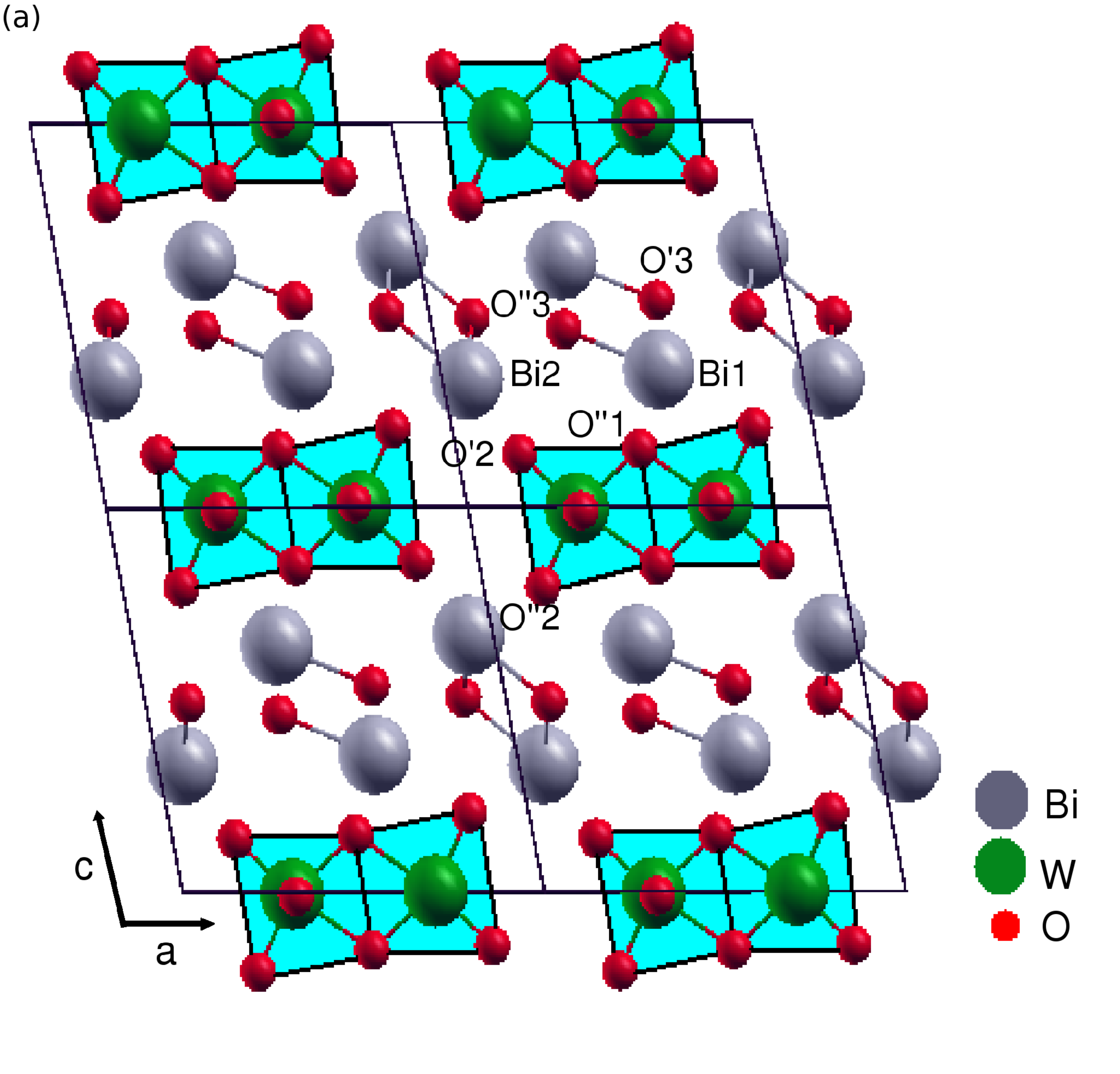}\\
\centering\includegraphics[angle=0,scale=0.090]{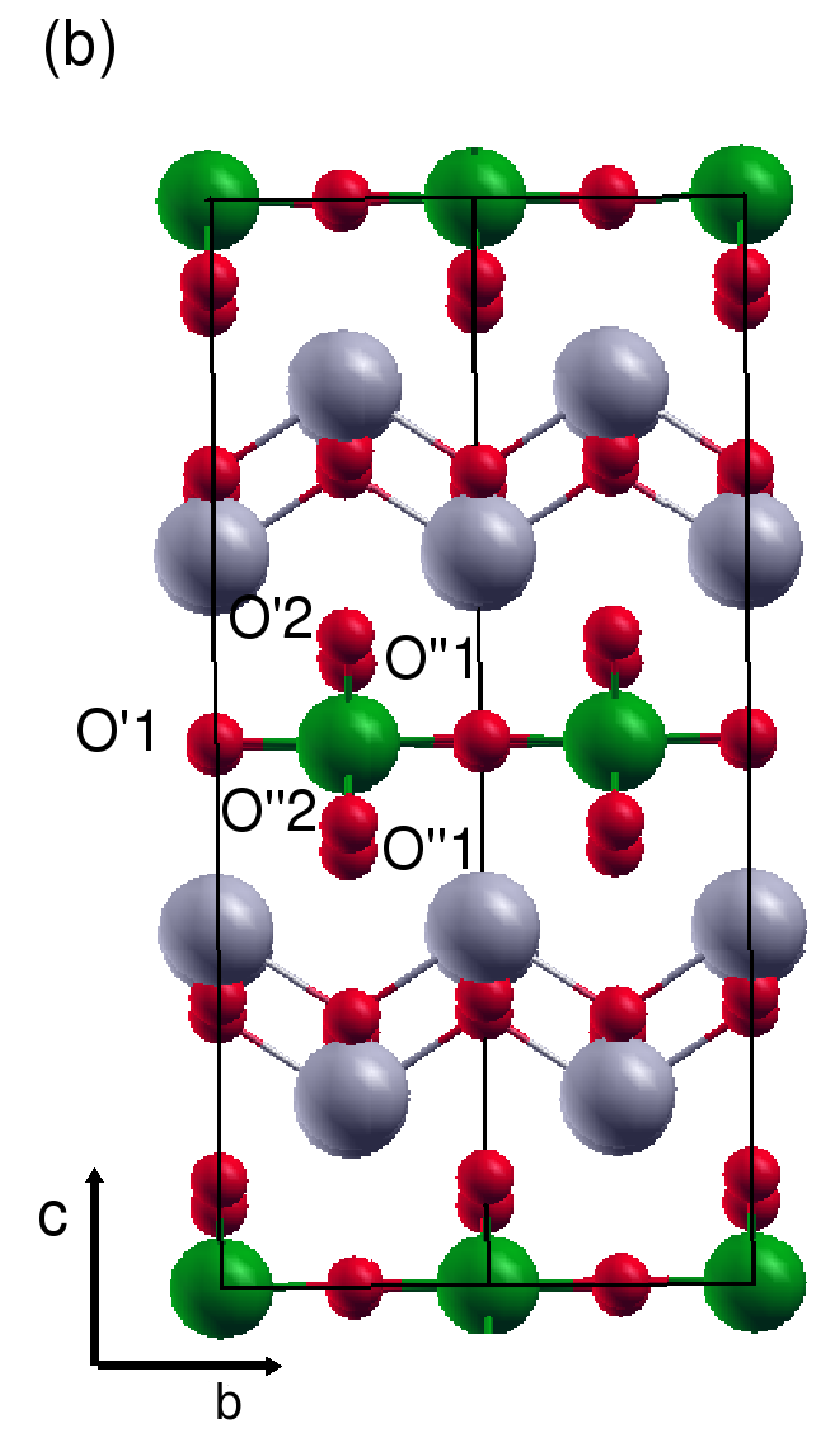}\\
\caption{\label{monoc} The monoclinic $A2/m$ paraelectric phase of Bi$_2$WO$_6$ showing edge-sharing octahedra in the $ac$ plane  (a) and  corner-sharing octahedra in the $bc$ plane (b)  .}
\end{figure}

\begingroup
\squeezetable
\begin{table}[h]
\begin{center}
\caption{\label{Bndshifttable} Calculated cell parameters and internal atomic positions in the paraelectric $A2/m$ phase of Bi$_2$WO$_6$ . Experimental values from Ref.\onlinecite{13}) are reported between parentheses for comparison.} 

\begin{tabular}{cccccccc}
\hline\hline
& Atom & & Wyckoff & & x & y & z \\
\hline
a=$8.11\mathring{A}$ & Bi$_1$ & & 4i & & 0.9307 & 0.0000 & 0.1588\\
\hspace{5pt}(8.37) & & & & & (0.9310 & 0.0000 & 0.1640)\\
& Bi$_2$ & & 4i & & 0.3905 & 0.0000 & 0.1756\\
b=$3.75\mathring{A}$ & & & & & (0.3950 & 0.0000 & 0.1768)\\
\hspace{5pt}(3.85) & W & & 4i & & 0.2979 & 0.5000 & 0.0000 \\
& & &  & & (0.3007 & 0.5000 & 0.0000)\\
c=$15.90\mathring{A}$& O'$_1$ & & 4i & & 0.3142 & 0.0000 & 0.0024\\
 \hspace{5pt}(16.44)& & & & & (0.3113 & 0.0000 & 0.0035)\\
 & O''$_1$ & & 4i & & 0.5090 & 0.5000 & -0.0811\\
$\beta$=$102.12\textdegree$& & & & & (0.5080 & 0.5000 & -0.0770)\\
 \hspace{9pt}(102.33)& O'$_2$ & & 4i & & 0.1638 & 0.5000 & -0.0736\\
 & & & & & (0.1734 & 0.5000 & -0.0710)\\
& O''$_2$ & & 4i & & 0.1661 & 0.5000 & 0.1059\\
& & & & & (0.1640 & 0.5000 & 0.1009)\\
& O'$_3$ & & 4i & & 0.1172 & 0.0000 & 0.2500\\
& & & & & (0.1170 & 0.0000 & 0.2512)\\
& O''$_3$ & & 4i & & 0.3618 & 0.5000 & 0.2660\\
& & & & & (0.3638 & 0.5000 & 0.2668)\\
\\
\hline\hline
\end{tabular}
\end{center}
\end{table}
\endgroup

\section{Technicalities}

The calculations were performed within density functional theory \cite{18, 19}, using a plane waves pseudopotentials method thanks to the ABINIT package \cite{20, 21, 22}. 
The exchange-correlation energy functional was evaluated within the Local Density Approximation (LDA) as parameterized by Perdew and Wang \cite{23}. Extended norm conserving pseudopotentials from Teter \cite{24} were used, with Bi(5d 6s 6p), W(5s 5p 5d 6s) and O(2s 2p) levels treated as valence states. The wave functions were expanded up to a kinetic energy cutoff of 50 Hartrees. Integrals over the Brillouin zone were approximated by sums on a 6x6x1 Monkhorst-Pack mesh of special k-points \cite{25}.
The structural optimization was done using the Broyden-Fletcher-Goldfarb-Shanno minimization algorithm (BFGS) \cite{26}.  We calculated the \textit{ab initio} forces on the ions and relaxed the position of each individual atom  until the absolute values of the forces were converged to less than 10$^{-5}$ Ha/Bohr. Phonons, Born effective charges and dielectric tensors were calculated using Density Functional Perturbation Theory \cite{27} and spontaneous polarization using the Berry phase formalism \cite{28}.

\section{High temperature $A2/m$ monoclinic structure}
We begin our study by fully characterizing the experimentally observed paraelectric $A2/m$ phase. In this structure for the first time, phonon instabilities are calculated together with dielectric properties and Born effective charges. Further, we will move on with the characterization of the hypothetical paraelectric $I4/mmm$ phase. 
\subsection{Structure}

The monoclinic $A2/m$ paraelectric phase of Bi$_2$WO$_6$ (conventional $C2/m$) contains 2 formula units per primitive unit cell (18 atoms). The structure, illustrated in Fig.1, consists of alternative Bi$_2$O$_2$ fluorite-like  and WO$_{4}$ layers. Within this structure the oxygen octahedra exhibit a peculiar edge-sharing along $a$-axis. Along $b$-axis, the oxygen octahedra are corner-sharing. As conventional for Aurivillius phases, the stacking axis normal to the layers is defined as the $c$-axis of the $A2/m$ structure. The calculated lattice parameters and internal positions are shown in Table I. Except for the systematic LDA underestimate of lattice parameters, the results are in reasonable agreement with experiment.  We notice that the lattice constants of Mohn and Stolen \cite{17} within the generalized gradient approximation (GGA) are in closer agreement with the experiment. This might however be partly fortuitous: at the opposite to LDA, GGA is well know to overestimate lattice constants and this typical error (observed in the lower-temperature $B2cb$ phase, as discussed in a next Section) might be partly compensated here by the thermal expansion, since the experimental results to which the zero-temperature calculations are compared have been done at 965$^{\circ}$C.

\subsection{Dielectric properties}
The study of the Born effective charges ($Z^*$) and optical dielectric tensor ($\varepsilon_\infty$) is important for the identification of the long-range dipolar contribution to the lattice dynamic of a polar insulator. Our results for $Z^*$ of the non-equivalent atoms are presented in Table II.  For clarity, we note that $A2/m$ has its $a$ and $b$ axis aligned along the cartesian $x$ and $y$ directions, respectively, and $c$-axis  makes an angle of ($\beta$ - $\frac{\pi}{2}$) with $z$ direction. Within this choice of cartesian axis, the Born effective charge tensor has the form:
\begin{eqnarray*}
Z^{*}= \left( \begin{array}{ccc} Z_{xx}^{*} & 0 & Z_{xz}^{*} \\0 & Z_{yy}^{*} & 0 \\ Z_{zx}^{*} & 0 & Z_{zz}^{*} \end{array} \right) 
\end{eqnarray*}

\begin{table}
\caption{Non-zero elements of the calculated Born effective charge tensors in cartesian coordinates (see text) for  Bi$_2$WO$_6$ in the monoclinic $A2/m$ phase. The nominal charge of each atom ($Z_{nom}$) is reported for comparison}
\begin{tabular}{lrrrrrr}
\hline\hline
Atom & $Z_{xx}^{*}$ & $Z_{yy}^{*}$ & $Z_{zz}^{*}$ & $Z_{xz}^{*}$ & $Z_{zx}^{*}$ & $Z_{nom}$ \\
\hline
Bi$_1$ & 5.34 & 5.38 & 4.35 & 0.24 & -0.36 & +3\\ 
Bi$_2$ & 4.37 & 5.17 & 4.53 & -0.18 & 0.18 & +3\\
W & 6.58 & 10.5 & 7.33 & 0.08 & 0.63 & +6\\
O'$_1$ & -1.21 & -7.76 & -1.34 & -0.01 & -0.06 & -2\\
O''$_1$ & -3.27 & -2.68 & -3.33 & -0.75 & -0.91 & -2\\
O'$_2$ & -2.73 & -2.18 & -2.71 & 2.00 & 2.02 & -2\\
O''$_2$ & -2.34 & -2.5 & -3.86 & -1.33 & -1.44 & -2\\
O'$_3$ & -3.25 & -2.97 & - 2.68 & -0.21 & 0.04 & -2\\
O''$_3$ & -3.46 & -2.96 & -2.27 & 0.16 & -0.10 & -2\\
\hline\hline
\end{tabular}
\end{table}

The Born effective charges for this material are much larger than the nominal ionic ones and so follow the tendency observed for the class of perovskite ABO$_3$ compounds \cite{29}. Giant anomalous Z$^*$ are observed on W (+10.5$e$) and O'$_1$ (-7.76$e$) belonging to the W-O'$_1$ chains aligned along $b$-axis.  These values are similar to those reported for WO$_3$ \cite{30} and can be assigned to W 5d -- O 2p dynamical changes of hybridization. Anomalous parts of the charge on W and on its surrounding  O''$_1$, O'$_2$, and O''$_2$ oxygen atoms, are strongly reduced along $x$ and $z$ directions where W-O chains are broken. Anomalous Z$^*$ are also observed in the Bi$_2$O$_2$ layer with a large contribution observed on Bi and its surrounding oxygen atoms. One can see a significant difference between Bi$_1$ and Bi$_2$ (in $x$ direction) despite their apparently similar environment in the Bi$_2$O$_2$ layer. The Born effective charge is known to be a sensitive tool to probe covalency effects  \cite{29} and our results therefore support the recent work of Mohn \textit{et al.} \cite{17}, who suggest that the local environment around Bi is asymmetric, with Bi$_1$ and Bi$_2$ forming different number of bonds with the apex oxygen of the perovskite layer.

Finally, the optical dielectric constant is computed:

\begin{eqnarray*}
\varepsilon_\infty= \left( \begin{array}{ccc} 6.49 & 0 & 0.08 \\0 & 7.25 & 0 \\ 0.08 & 0 & 6.37 \end{array} \right) 
\end{eqnarray*}
\\
We have no experimental values to compare with and our prediction has to be considered as an upper bound, because of the typical LDA overestimate of $\varepsilon_\infty$.

\begingroup
\squeezetable
\begin{table*}
\caption{Real-space eigendisplacements $\eta$ of the unstable phonon modes of the $A2/m$ structure. The eigendisplacements are normalized within the 36 atoms cell according to  $\textless \eta \textbar M \textbar \eta \textgreater = 1$,  with the mass matrix M in atomic mass unit \cite{note}. Internal coordinates of the atoms of the $A2/m$ phase are those presented in Table I. Modes frequencies between brackets are in cm$^{-1}$.}
\begin{tabular}{lrrrrrrrrrrrrrrrrrrrrrrrrrrrrrrrrrr}
\hline\hline
Atom & & & & & & & & & & & & & $\eta$ & & & & & & & & & & & & & &\\
\hline
& & & & & $\Gamma_1^{-}$ & & & & & & & $V_1^{-}$ & & & & & & &  $V_1^{+}$ & & & & & & &  $\Gamma_2^{+}$ & &\\
& & & & & [108$i$] & & & & & & & [89$i$] & & & & & & &  [65$i$] & & & & & & &  [63$i$] & &\\
& & & & x & y & z & & & & & x & y & z & & & & & x & y & z & & & & & x & y & z &\\
Bi$_1$ & & & & 0.000 & -0.001 & 0.000 & & & & & 0.000 & 0.006 & 0.000 & & & & & 0.000 & 0.004 & 0.000 & & & & & 0.000 & 0.003 & 0.000 &\\ 
Bi$_2$ & & & & 0.000 & 0.000 & 0.000 & & & & & 0.000 & 0.004 & 0.000 & & & & & 0.000 & 0.007 & 0.000 & & & & & 0.000 & 0.004 & 0.000 &\\
W & & & & 0.000 & 0.021 & 0.000 & & & & & 0.000 & 0.027 & 0.000 & & & & & 0.000 & 0.035 & 0.000 & & & & & 0.000 & 0.036 & 0.000 &\\
O'$_1$ & & & & 0.000 & -0.032 & 0.000 & & & & & 0.000 & -0.023 & 0.000 & & & & & 0.000 & 0.002 & 0.000 & & & & &  0.000 & 0.003 & 0.000 &\\
O''$_1$ & & & & 0.000 & -0.045 & 0.000 & & & & & 0.000 & -0.037 & 0.000 & & & & & 0.000 & -0.012 & 0.000 & & & & &  0.000 & -0.007 & 0.000 &\\
O'$_2$ & & & & 0.000 & -0.044 & 0.000 & & & & & 0.000 & -0.034 & 0.000 & & & & & 0.000 & -0.005 & 0.000 & & & & & 0.000 & -0.022 & 0.000 &\\
O''$_2$ & & & & 0.000 & -0.067 & 0.000 & & & & & 0.000 & -0.055 & 0.000 & & & & & 0.000 & -0.024 & 0.000 & & & & & 0.000 & -0.022 & 0.000 &\\
O'$_3$ & & & & 0.000 & -0.022 & 0.000 & & & & & 0.000 & 0.002 & 0.000 & & & & & 0.000 & -0.002 & 0.000 & & & & & 0.000 & -0.002 & 0.000 & \\
O''$_3$ & & & & 0.000 & -0.020 & 0.000 & & & & & 0.000 & -0.012 & 0.000 & & & & & 0.000 & 0.006 & 0.000 & & & & & 0.000 & 0.001 & 0.000 & \\
\\
 \hline\hline
\end{tabular}
\end{table*}
\endgroup
\begin{table}[h]
\begin{center}
\caption{Cell parameters (lengths in \AA \,\, and angles in degrees)  and difference of energies $\Delta E_{m}$ with respect to the paraelectric monoclinic $A/2m$ phase (in meV/formula unit) after full atomic relaxation of the different phases reached by the condensation of individual unstable phonon modes in the $A/2m$ phase. The label and frequency $\omega$ ( in cm$^{-1}$) of the mode that was initially condensed, together with the symmetry of the related phase are also mentioned.}
\begin{tabular}{cccccc}
\hline\hline
 Mode&$\omega$&Symmetry &Cell parameters&& $\Delta E_{m}$  \\
\hline
$\Gamma_1^{-}$&108$i$&$A$2 & {\scriptsize a=8.09  b=3.77 c=15.87} && $-7.82$ \\
                                  &(polar) & &{\scriptsize $\alpha$=90\textdegree  $\beta$=101.50\textdegree $\gamma$= 90\textdegree} &&\\
\\                                                                   
$V_1^{-}$ & 89$i$ & $P\overline{1}$ & {\scriptsize a=8.09  b=3.77 c=15.88} && $-6.87$ \\
 & & &{\scriptsize $\alpha$=90.00\textdegree $\beta$=101.50\textdegree  $\gamma$= 89.99\textdegree} &&\\
\\
$V_1^{+}$  & 65$i$ & $P\overline{1}$  & {\scriptsize a=8.09  b=3.77 c=15.87} && $-9.25$ \\
& & &{\scriptsize $\alpha$=89.99\textdegree  $\beta$=101.50\textdegree  $\gamma$= 89.99\textdegree} &&\\
\\
$\Gamma_2^{+}$  & 63$i$ & $P\overline{1}$  & {\scriptsize a=8.09  b=3.77 c=15.87} && $-8.16$ \\
& & &{\scriptsize $\alpha$=90.30\textdegree  $\beta$=101.50\textdegree  $\gamma$= 90.10\textdegree} &&\\

\hline\hline
\end{tabular}
\end{center}
\end{table} 

\subsection{Phonon instabilities}

Phonon calculations were performed to study the structural stability of the $A2/m$ monoclinic configuration. The zone-center phonons were computed in the conventional cell of 36 atoms, with the primitive cell doubled along the $b$-axis. Due to the folding of the Brillouin zone, this gives access to the phonons at $\Gamma$ and $V$. The full list of $\Gamma$ frequencies is reported in Appendix for completeness. The system reveals four instable modes : two at $\Gamma$ with frequencies at 108$i$ cm$^{-1}$ and 63$i$ cm$^{-1}$, corresponding to the irreducible representations labels \cite{31} $\Gamma_1^{-}$ and  $\Gamma_2^{+}$, respectively, and two others at the zone-boundary point $V$=(0, 1/2, 0),  with frequencies at 89$i$ and 65$i$ cm$^{-1}$, corresponding to $V_1^{-}$ and $V_1^{+}$, respectively. The strongest unstable mode $\Gamma_1^{-}$ is polar along  $b$-axis.
From Table III, where real-space eigendisplacements of these unstable modes are given, we can see that all of them restrict their atomic motions along $b$-axis direction only and that contrary to what is usually observed in other paraelectric Aurivillius phases, the $A2/m$ monoclinic configuration is not unstable against oxygen octahedra rotational modes, which are prevented by the edge-sharing spatial organization.

The unstable modes were then individually condensed into the $A2/m$ phase and each of the lower symmetry phases reached under these condensations is fully relaxed. From the results given in Table IV, we can see that the condensation of the polar unstable mode $\Gamma_1^{-}$ into $A2/m$ leads to a ferroelectric phase with monoclinic $A$2 (No. 5) structure, while the condensation of V$_1^{-}$, V$_1^{+}$ and $\Gamma_2^{+}$ lead to monoclinic $P\overline{1}$ (No. 2) structure. After the full relaxation, these phases exhibit nearly identical cell parameters and total energy values close to each others. The lowest energy configuration corresponds to the  monoclinic $P\overline{1}$  phase arising from the condensation of V$_1^{+}$ with a decrease of energy, with respect to $A2/m$ phase, of 9.25 meV/formula unit. Further phonon calculations in this phase do not reveal any unstable mode, identifying it as a potential ground state.

\subsection{Spontaneous polarization}

As highlighted in the previous section, starting from the $A2/m$ phase, a ferroelectric $A$2  structure can be reached from the condensation of an unstable $\Gamma_1^{-}$ mode, giving rise to a polarization along the $b$ direction. The amplitude of this polarization can be computed using the Berry phase formalism but requires to be cautious. 

Indeed, the modern theory of the polarization (see for instance Ref. \cite{28}) teaches us that the spontaneous polarization $P_s$ of the ferroelectric $A$2  structure is related to the {\it formal} polarizations of the ferroelectric ($P_{A2}$) and paraelectric ($P_{A2/m}$) phases by 
\begin{equation}
P_s = P_{A2} - P_{A2/m} + n \textbar P_Q \textbar    
 \label{Eq-Ps}
\end{equation}
where $n$ is an integer and $P_Q$ = 2 e $\bf{R}/\Omega$ is the so-called polarization quantum, with $\bf{R}$ the shortest lattice vector along the polarization direction and $\Omega$ the unit-cell volume. The factor of 2 accounts for the double band occupancy. 

From Eq. \ref{Eq-Ps}, the determination of $P_s$ from the only knowledge of the formal polarizations $P_{A2}$ and $P_{A2/m}$ is only possible modulo $P_Q$. This quantum indeterminacy arises from the fact that the formal polarizations of the $P_{A2}$ and $P_{A2/m}$ phases are themselves only defined modulo $P_Q$ so that the value of $P_s$ depends on the specific choice of a polarization branch (i.e. choice of how we connect a given $P_{A2/m}$ point on the left to a given $P_{A2}$ point on the right in Figure 2). Of course, $P_s$ is a well-defined quantity but its unambiguous determination additionally requires to determine $n$ through the proper integration of the polarization current along an adiabatic insulating path between the paraelectric and ferroelectric phases.

When the polarization quantum is large compared to the spontaneous polarization, like in conventional ABO$_3$ compounds, the uncertainty can be easily fixed. But in materials like Aurivillius compounds that have large unit cells, the quantum is small and potentially of the same order of magnitude as the spontaneous polarization (in the present case $P_Q$=25.90 $\mu$C/cm$^2$) and getting rid of the quantum indeterminacy requires to follow the evolution of the polarization along the atomic distortion from the paraelectric to the ferroelectric phase.  In practice, this can be achieved by computing the formal polarizations of intermediate structures as illustrated in Figure 2.  The choice of the appropriate branch (solid lines in the graph) becomes obvious when the change of polarization from one intermediate structure to the next one is small compared to the polarization quantum. From this, we deduce that the spontaneous polarization of the $A2$ phase is $P_s = 20.02$ $\mu$C/cm$^2$ . We notice that this value differs significantly from the one that would have been naively deduced from the difference between the values of the formal polarizations in the interval $[-P_Q/2, P_Q/2 ]$, as typically delivered in the output of conventional first-principles softwares (i.e. $P_{A2}$= -5.88$\mu$C/cm$^2$ and $P_{A2/m}$=0$\mu$C/cm$^2$ yielding $P_s$ = -5.88$\mu$C/cm$^2$).

Alternatively, it is also possible to estimate the polarization from the knowledge of the Born effective charges and atomic displacements using the expression: 
\begin{equation}
P_{s,\alpha} = \frac{e}{\Omega}\sum_{\kappa, \beta}Z_{\kappa,\alpha\beta}^{*}\delta\tau_{\kappa,\beta},
\label{Eq-Z}
\end{equation}   
where  $\delta\tau_{\kappa,\beta}$ is the displacement of atom $\kappa$ along direction $\beta$ from the paraelectric to the ferroelectric phase and $Z_{\kappa,\alpha\beta}^{*}$ the Born effective charge tensor of atom $\kappa$.  This provides $P_s=20.73$ $\mu$C/cm$^2$ in close agreement with the Berry phase estimate and coherent with the quasi-linear evolution of the polarization with the structural distortion seen in Figure 2.  We notice that Equation 2 gives access to the slope at the origin of the appropriate polarization branch in Figure 2 and provides an alternative easy way to identify it. 

\begin{figure}[t]
\centering\includegraphics[angle=0,scale=0.36]{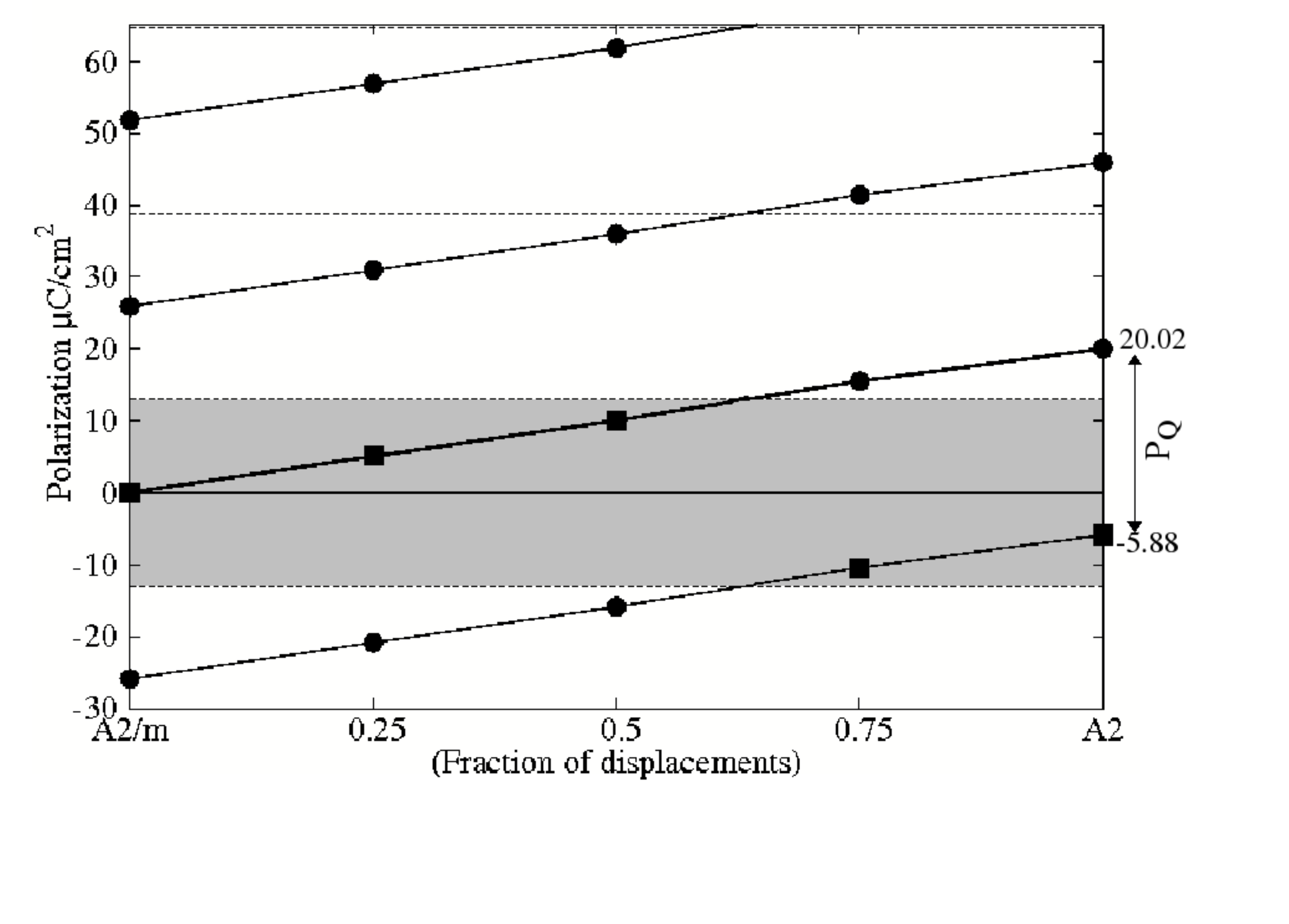}
   \caption{\label{pola} Change in polarization P along a path of atomic distortions from the paraelectric $A2/m$ structure to the ferroelectric $A2$ structure. The possible values of P at given atomic positions differ by multiples of the polarization quantum, $P_Q$=25.90 $\mu$C/cm$^2$. Squares represent formal polarizations in the interval $[-P_Q/2, P_Q/2 ]$(grey area). }  
\end{figure}

\begin{figure}[t]
  \centering\includegraphics[angle=0,scale=0.090]{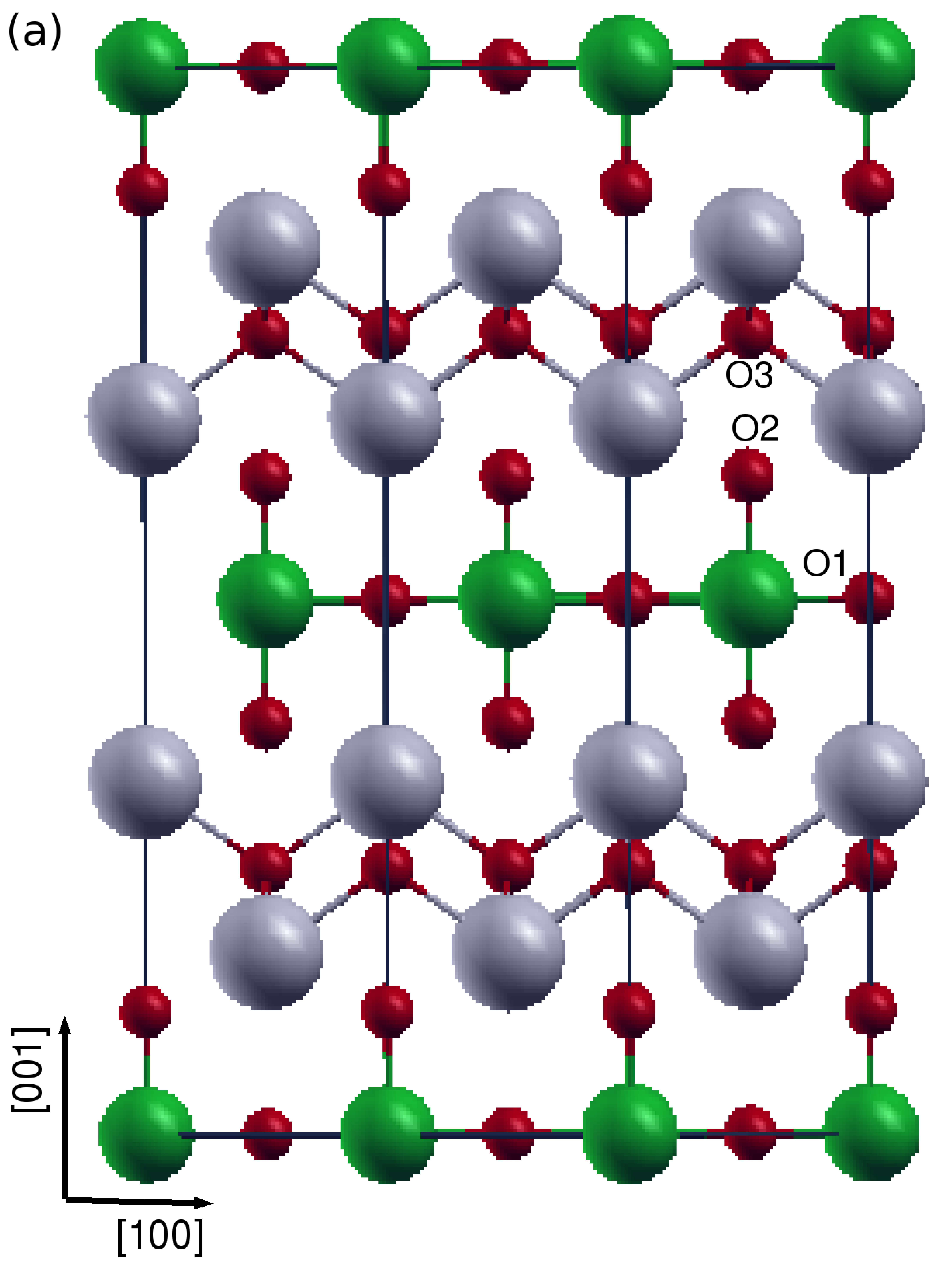}
  \centering\includegraphics[angle=0,scale=0.080]{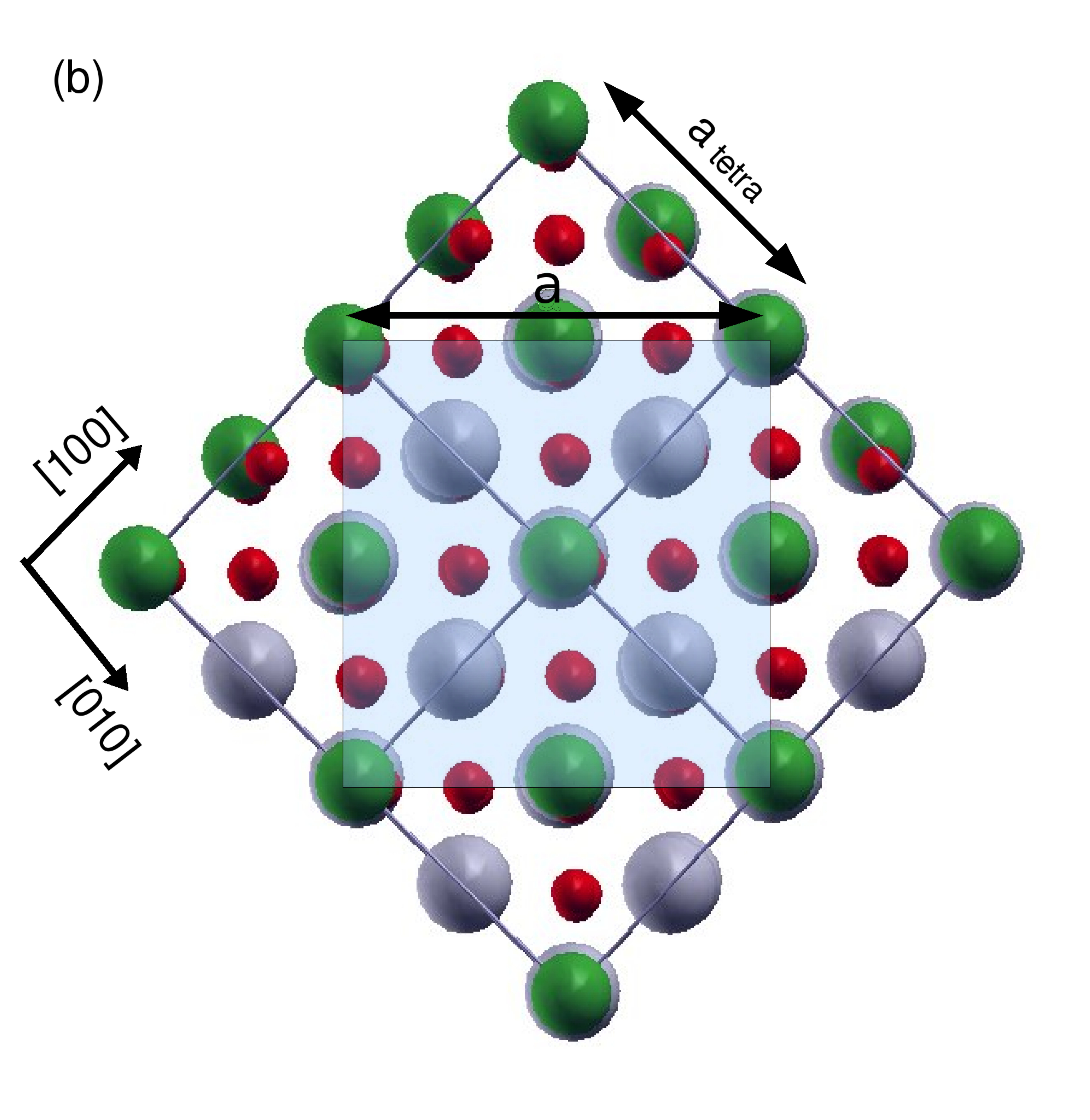}\\
   \caption{\label{tetradoubl} The hypothetical tetragonal $I4/mmm$ paraelectric  phase of Bi$_2$WO$_6$: side view (a) and top view (b), highlighting the corresponding doubled cell in [110] direction.}
\end{figure}
\begingroup
\squeezetable
\begin{table}[h]
\begin{center}
\caption{\label{Bndshifttable} Calculated unit cell parameters and internal atomic positions in the hypothetical paraelectric $I4/mmm$ phase doubled along [110] and the $B2cb$ and $P2_1ab$ ferroelectric phases. Experimental values estimated from ref.\cite{16} for the tetragonal paraelectric phase (see text) and taken from ref.\cite{13} for orthorhombic ferroelectric phases are reported in parentheses for comparison. We are not reporting Wyckoff positions for the $I4/mmm$ phase since our unit cell is doubled. }
\begin{tabular}{ccccccccc}
\hline\hline
Structure& & Atom & & Wyckoff & & x & y & z \\
\hline
$I4/mmm$   & & & & & & & & \\
\\
a=$5.31\mathring{A}$ & & Bi & & & & 0.0000 & 0.5000 & 0.1756\\
 \hspace{5pt}($5.45$) &  & & & & & (0.0000 & 0.5000 & 0.1701) \\ 
  &  & W & &  & & 0.0000 & 0.0000 & 0.0000\\ 
 c=$16.14\mathring{A}$ &  & O$_1$ & & & & 0.2500 & 0.7500 & 0.0000\\
 \hspace{5pt}($16.52$)& &  O$_2$ & & & & 0.0000 & 0.0000 & 0.1143\\
 & & & & & & (0.0000 & 0.0000 & 0.1152)\\
 & & O$_3$ & &  & & 0.2500 & 0.2500 & 0.7500\\
\\
\hline
$B2cb$  & & & & & & & &\\
\\ 
a=$5.35\mathring{A}$ & & Bi & & 8b & & 0.9961 & 0.5149 & 0.1702\\
 \hspace{5pt}($5.49$) &  & & & & & (0.9947 & 0.5116 & 0.1722)\\
 &  & W & & 4a & & 0.0000 & 0.0000 & 0.0000\\
b=$5.38\mathring{A}$ & & O$_1$ & & 8b & & 0.2173 & 0.7534 & 0.9828\\
 \hspace{5pt}($5.52$) & & & & & & (0.2981 & 0.7399 & 0.9892)\\
& & O$_2$ & & 8b & & 0.0641 & 0.9536 & 0.1101\\
c=$16.12\mathring{A}$ & & & & & & (0.0753 & 0.9587 & 0.1087)\\
\hspace{5pt}($16.54$) & & O$_3$ & & 8b & & 0.2369 & 0.2547 & 0.7484\\
& & & & & & (0.2771 & 0.2550 & 0.7499)\\
\\
\hline
$P2_{1}ab$  & & & & & & & &\\
\\
$a=5.30\mathring{A}$ & & Bi$_1$ & & 4a & & 0.0109 & 0.4843 & 0.1698\\
\hspace{5pt}(5.45) &  & & & & & (-0.0126 & 0.5113 & 0.1730)\\
 &  & Bi$_2$ & & 4a & & 0.0036 & 0.5074 & -0.1698\\
b=$5.32\mathring{A}$ & & & & & & (-0.0113 & 0.4762 & -0.1719)\\
 \hspace{5pt}(5.48)& & W & & 4a & & 0.0000 & 0.0000 & 0.0000\\
 & & O'$_1$ & & 4a & & 0.2318 & 0.6896 & 0.0108\\
c=$16.17\mathring{A}$ & & & & & & (0.2679 & 0.6933 & -0.0147)\\
\hspace{5pt}(16.47) & & O''$_1$ & & 4a & & 0.3321 & 0.1984 & -0.0109\\
& & & & & & (0.3342 & 0.2220 & 0.0163)\\
& & O'$_2$ & & 4a & & -0.0684 & -0.0368 & 0.1109\\
& & & & & & (0.0854 & 0.0449 & 0.1072)\\
& & O''$_2$ & & 4a & & 0.4428 & 0.4563 & -0.1108\\
& & & & & & (0.5703 & 0.5526 & -0.1086)\\
& & O'$_3$ & & 4a & & 0.2676 & 0.2413 & 0.7515\\
& & & & & & (0.2728 & 0.2508 & 0.7489)\\
& & O''$_3$ & & 4a & & 0.2752 & 0.2465 & 0.2516\\
& & & & & & (0.2740 & 0.2326 & 0.2514)\\
\\
\hline\hline
\end{tabular}
\end{center}
\end{table}
\endgroup

\section{Hypothetical high temperature $I4/mmm$ tetragonal structure}

Starting at high-temperature from the experimentally observed paraelectric $A2/m$ structure, one would anticipate from the previous discussion, a transition to a non-ferroelectric $P\overline{1}$ ground state. Instead of that, experiments show that Bi$_2$WO$_6$ exhibits low temperature phases of orthorhombic $B2cb$ and  $P2_1ab$ symmetries,  subgroups of the centrosymmetric $I4/mmm$ parent phase, that is the high temperature paraelectric phase of all Aurivillius compounds with a number of perovskite blocks m>1. We so continue our study by considering the hypothetical $I4/mmm$ paraelectric phase. 

\subsection{Structure}

The $I4/mmm$ structure as illustrated in Fig.3(a), contains 2 formula units per primitive unit cell (18 atoms) and consists of Bi$_2$O$_2$ fluorite-like layers alternating with perovskite-like blocks of corner-sharing octahedra chains. Since $I4/mmm$ is not observed experimentally, an estimate of the paraelectric unit cell is deduced from the orthorhombic lattice parameters of the experimental ferroelectric phase cited in Ref. \cite{16}, using  a$_{tetra}$ = b$_{tetra}$ = 3.86$\mathring{A}$ (= $\sqrt{a_{orth} b_{orth}/2 }$)  and c$_{tetra}$ = 16.52$\mathring{A}$ . In order to facilitate the comparison with the experimentally observed orthorhombic ferroelectric ground state, that contains 4 formula units per primitive unit cell (36 atoms), our calculations were performed in a tetragonal $I4/mmm$ unit cell doubled along the [110]  directions, where the lattice parameters become \textit{a} = \textit{b} = 5.45 $\mathring{A}$, \textit{c} = 16.52$\mathring{A}$ (see Fig.3 (b)),  allowing the paraelectric and ferroelectric phases to have the same number of atoms and the same orientation. In what follows, a,b and c will be the vectors of our doubled unit cell so that $a$ and $b$ directions will refer to the [110] directions of the conventional centered unit cell and $c$ to the direction perpendicular to them.  

\textcolor{black}{The structural parameters and internal atomic positions of the $I4/mmm$ hypothetical structure after full atomic relaxation are reported in the top part of Table V. Our internal coordinates show good agreement with those calculated by Machado {\it et al.} in Ref.\cite{16} who worked however at larger lattice constants, corresponding to experimental values estimated as discussed above. } 

We further notice that the internal energy of the relaxed $I4/mmm$ structure is 156 meV per formula unit  higher than that of the $A2/m$ phase. From now, we will report the energies taking the $I4/mmm$ structure as reference. A summary of the internal energies of all the phases studied will be provided in Section V.

\subsection{Dielectric Properties}

Our results for the dynamical effective charges of Bi$_2$WO$_6$ in the $I4/mmm$ phase, for non-equivalent atoms are presented in Table VI. In this case, the cartesian $x$, $y$, and $z$ axis are aligned along the crystalline  $a$, $b$ and $c$ directions.  

\begin{table}[h]
\begin{center}
\caption{\label{Bndshifttable}Non-zero elements of the calculated Born effective charges tensors in cartesian coordinates for Bi$_2$WO$_6$ in the hypothetical tetragonal $I4/mmm$ symmetry.}
\begin{tabular}{lrrrr}
\hline\hline
Atom & $Z_{xx}^{*}$ & $Z_{yy}^{*}$ & $Z_{zz}^{*}$ & $Z_{nom}$\\
\hline
Bi & 5.00 & 5.00 & 4.29 & +3 \\
W  & 11.09 & 11.09 & 9.32 & +6 \\
O$_1$ & -4.88 & -4.88 & -1.46 & -2 \\
O$_2$ & -2.7 & -2.7 & -4.74 & -2 \\
O$_3$ & -2.95 & -2.95 & -2.74 & -2 \\
\hline\hline
\end{tabular}
\end{center}
\end{table} 

As expected, there are large anomalous contributions to $Z^{*}$, on  W(+11$e$) and O$_1$(-4.88$e$) atoms in the octahedral $ab$-plane and on W(+9.32$e$) and O$_2$(-4.74$e$) atoms in the apical direction. Within  Bi$_2$O$_2$ layers, the large charges also observed on Bi(+5$e$) attest of the ionico-covalent character for bonding in both the fluorite-like and perovskite-like units of Aurivillius structures. Comparison with $A2/m$ shows that $Z^{*}$ on W along W-O chains are in the same range of values for both structures, but O'$_1$ (-7.76$e$) in $A2/m$ is much larger than O$_1$(-4.88$e$) in $I4/mmm$.

Optical dielectric tensor $\varepsilon_\infty$  is given by:
\begin{eqnarray*}
\varepsilon_\infty= \left( \begin{array}{ccc} 7.60 & 0 & 0 \\0 & 7.60 & 0 \\ 0 & 0 & 6.84 \end{array} \right) 
\end {eqnarray*}
These values are comparable to those reported for the $A2/m$ phase.

\subsection{Phonons Instabilities}
 \textcolor{black}{Starting from fully relaxed tetragonal prototype structure, we computed the zone-center phonon modes of the doubled 36 atoms unit cell that, due to the folding of the Brillouin zone, combine in fact phonons at the $\Gamma$ and $X$ points of the Brillouin zone of the primitive unit cell.} We identified five unstable phonon modes, twofold degenerated along $a$ and $b$ directions. The mode with the largest instability lies at the Brillouin zone center $\Gamma$ with a frequency at 198$i$ cm$^{-1}$ and corresponding to the irreducible representation $\Gamma_5 ^{-}$. The four other unstable modes are at the zone boundary point $X$=($\frac{1}{2}$, $\frac{1}{2}$, 0) \cite{32}  with  frequencies at 183$i$ ,135$i$, 104$i$ and 98$i$ cm$^{-1}$, and corresponding to the irreducible representations, $X_2^{-}$ ,$X_2^{+}$, $X_3^{+}$ and $X_4^{+}$, respectively. The latter four modes are antiferrodistortive modes since they are at the zone boundary $X$; their condensation in the tetragonal structure will necessarily lead to the unit cell doubling along the [110] direction. \textcolor{black}{The complete list of $\Gamma$ phonons and a comparison with the results of Machado {\it et al.} \cite{16} is provided in Appendix}.

\begingroup
\squeezetable
\begin{table*}
\caption{Real-space eigendisplacements $\eta$ of unstable and some relevant stable phonon modes of the hypothetical $I4/mmm$ structure. The eigendisplacements  are normalized for the 36 atoms cell according to $\textless \eta \textbar M \textbar \eta \textgreater = 1$,  with the mass matrix M in atomic mass unit \cite{note}. The internal coordinates of the different atoms are those of Table V. The phonon frequencies, reported in brackets, are in cm$^{-1}$.}
\begin{tabular}{crrrrrrrrrrrrrrrrrrrrrrrrrrrrrr}
\hline\hline
Atom & & & & & & & & & & & & & $\eta$ & & & & & & & & & & & & & &&\\
\hline
& & & & & $\Gamma_5^{-}$(a) & & & & & & & $X_2^{-}$(a) & & & & & & & $X_2^{+}$(a) & & & & & & & & &\\
& & & & & [198$i$] & & & & & & & [183$i$] & & & & & & & [135$i$] & & & & & & & & &\\
& & & & x & y & z & & & & & x & y & z & & & & & x & y & z & & & & &  &  &  &\\
Bi & & & & 0.0020 & 0.0000 & 0.0000 & & & & & 0.0040 & 0.0000 & 0.0000 & & & & & 0.0000 & 0.0000 & 0.0000 & & & & & & & &\\ 
W & & & & 0.0170 & 0.0000 & 0.0000 & & & & & 0.0190 & 0.0000 & 0.0000 & & & & & 0.0000 & 0.0000 & 0.0000 & & & & &  &  &  &\\
O$_1$ & & & & -0.0400 & -0.0020 & 0.0000 & & & & & -0.0430 & -0.0050 & 0.0000 & & & & &-0.0620 & -0.0620 & 0.0000 & & & & & & & &\\ 
O$_2$ & & & & -0.0660 & 0.0000 & 0.0000 & & & & & -0.0610 & 0.0000 & 0.0000 & & & & & 0.0000 & 0.0000 & 0.0000 & & & & &  &  &  &\\
O$_3$ & & & & -0.0150 & 0.0000 & 0.0000 & & & & & 0.0000 & -0.0010 & 0.0000 & & & & & 0.0002 & 0.0000 & 0.0000 & & & & &  &  &  & \\
\\
\hline
& & & & & $X_3^{+}$(a) & & & & & & & $X_4^{+}$(a) & & & & & & & $\Gamma_5^{-}$ & & & & & & & & &\\
& & & & & [104$i$] & & & & & & & [98$i$] & & & & & & & [29] & & & & & & & & &\\
& & & & x & y & z & & & & & x & y & z & & & & & x & y & z & & & & &  &  &  &\\
Bi& & & & 0.0000 & 0.0000 & 0.0000 & & & & & 0.0000 & 0.0060 & 0.0000 & & & &  & 0.0150 & 0.0000 & 0.0000 & & & & & &  &  &\\
W & & & & 0.0000 & 0.0000 & 0.0000 & & & & & 0.0000 & 0.0000 & 0.0000 & & & &  & -0.0270 & 0.0000 & 0.0000 & & & & &  &  &  &\\
O$_1$  & & & & 0.0000 & 0.0000 & 0.0470 & & & & &  0.0000 & 0.0000 & -0.0510 & & & &  & -0.0230 & -0.0010 & 0.0000 & & & &  &  & &  & \\ 
O$_2$  & & & & 0.0000 & 0.0640 & 0.0000 & & & & & 0.0000 & 0.0690 & 0.0000 & & & &  & -0.0200 & 0.0000 & 0.0000 & & & & &  &  &  &\\ 
O$_3$  & & & & 0.0000 & 0.0000 & -0.0060 & & & & & 0.0000 & 0.0000 & 0.0000 & & & &  & 0.0010 & 0.0000 & 0.0000 & & & & &  &  &  & \\
\\
\hline
& & & & & $X_2^{-}$ & & & & & & & $X_3^{+}$ & & & & & & &  $\Gamma_5^{-}$ & & & & & & & & & \\
& & & & & [80] & & & & & & & [107] & & & & & & &  [137] & & & & & & & & & \\
& & & & x & y & z & & & & & x & y & z & & & & & x & y & z & & & & &  &  & &\\
Bi & & & & 0.0220 & 0.0000 & 0.0000 & & & &  & 0.0000 & 0.0220 & 0.0000 & & & & &  0.0010 & 0.0000 & 0.0000 & & & & & & & & \\
W  & & & & 0.0050 & 0.0000 & 0.0000 & & & &  & 0.0000 & 0.0000 & 0.0000 & & & & &      -0.0040 & 0.0000 & 0.0000 & & & & & & & &\\
O$_1$   & & & &  -0.0300 & 0.0110 & 0.0000 & & & &  & 0.0000 & 0.0000 & 0.0270 & & & &  &   0.0630 & 0.0420 & 0.0000 & & & & & & & &\\ 
O$_2$   & & & & 0.0070 & 0.0000 & 0.0000 & & & &  & 0.0000 & 0.0300 & 0.0000 & & & & &    -0.0420 & 0.0000 & 0.0000 & & & & & & & &\\ 
O$_3$  & & & & 0.0000 & -0.0050 & 0.0000 & & & &  & 0.0000 & 0.0000 & -0.0080 & & & & &    -0.0130 & 0.0000 & 0.0000 &  & & & & & & &\\
\\
\hline\hline
\end{tabular}

\end{table*}
\endgroup

In what follows, we will describe the atomic motions associated to the different unstable modes as deduced from the inspection of the real-space eigendisplacements are reported in Table VII.

The strongest unstable mode $\Gamma_5 ^{-}$ ($\omega = 198i$ cm$^{-1}$), degenerated into $\Gamma_5^{-}$(a) and  $\Gamma_5^{-}$(b), corresponds to the polar motion of W and Bi atoms against the oxygens along the $a$-axis for $\Gamma_5^{-}$(a) or along $b$-axis for $\Gamma_5^{-}$(b), together with a small additional shift of O$_1$ octahedral oxygen atoms in the $ab$-plane.  

We notice that, in our calculations, only one polar mode is unstable, while the presence of a second unstable polar mode, related to a nearly rigid motion of Bi$_2$O$_2$ layer relative to the perovskite-like block (rigid-layer mode), was previously reported in Ref. \onlinecite{16}. Such a polar mode exists in our phonon calculations but appears at a slightly stable frequency of 29 cm$^{-1}$ (labeled as $\Gamma_5^{-}$[29] in Table VII.). 

This difference probably arises from the slightly different cell parameters used in both calculations.

\begin{table*}
\caption{Differences of energies with respect to the hypothetical paraelectric tetragonal $I4/mmm$ phase, $\Delta E_{t}$ (in meV/formula unit), after full structural relaxations of the different phases arising from the freezing of one or more unstable modes. The space group of each phase is given in non conventional setting with \textit{c} set as the long axis. The labels and frequencies $\omega$ (in cm$^{-1}$) of relevant phonon modes are also provided.}
\begin{tabular}{lccccccccccccccr}
\hline\hline
 Mode label &  & $\omega$ & & &  &Symmetry & &  &&  &&  &&  $\Delta E_{t}$  \\
\hline
Phases arising from single mode condensation  & & & &&&&&&& \\
$\Gamma_5^{-}$(a)  &  & 198$i$& &&  &$Fmm2$  && && & & & &-114.75 \\
$X_2^{-}$(a)  &  & 183$i$& &&  &$Amma$ & &&&&&&& -92.52 \\
$X_2^{+}$(a)  &  & 135$i$& &&  &$Abam$  &&&&&&&  & -99.63 \\
$X_3^{+}$(a)  &  & 104$i$& &&  &$Bmab$  &&&&&&&  & -71.01 \\
$X_4^{+}$(a)  & & 98$i$& &&  &$Amaa$  &  &&&&&&& -50.04 \\
\\
Experimentally observed phases  & & & &&&&&&& \\
$\Gamma_5^{-}$(a)+ $X_3^{+}$(a) &  & & &&  &$B2cb$  &  &&&&&&& -187.02 \\
$\Gamma_5^{-}$(a) + $X_3^{+}$(a) + $X_2^{+}$(b) &  & & &&  &$P2_{1}ab$  &  &&&&&&& -216.27 \\
\\
Hypothetical phases  & & & & &&&&&&\\
$\Gamma_5^{-}$(a) + $X_2^{+}$(b) &  & & &&  &$A2_1ma$  &  &&&&&&& -206.11 \\
$X^{3+}$(a) +  $X^{2+}$(b) &  & & &&  &$P2_1/a$  &  &&&&&&& -105.43 \\
$\Gamma ^{5-}$($a$) +$X^{3+}$($a$) +$X^{2+}$($a$) &  & & &&  &$Pa$  &  &&&&&&&  -213.05 \\
$\Gamma ^{5-}$($a$) +$X^{3+}$($b$) +$X^{2+}$($b$) &  & & &&  &$P2_1$  &  &&&&&&&  -204.21 \\

\hline\hline
\end{tabular}
\end{table*} 

The second unstable mode $X_2^{-}$ ($\omega = 183i$ cm$^{-1}$), degenerated into $X_2^{-}$(a) and $X_2^{-}$(b), corresponds to an anti-polar motion with displacements of  the atoms very similar to the $\Gamma_5^{-}$ mode, except for O$_3$ atoms.

The third unstable mode $X_2^{+}$  ($\omega = 135i$ cm$^{-1}$), degenerated into $X_2^{+}$(a) and $X_2^{+}$(b), consists in a rotation of oxygen octahedra about the $c$-axis. In this mode, W and O$_2$ atoms have no displacements while O$_1$ atoms shift in the $ab$-plane. 
The difference between  $X_2^{+}$(a) and $X_2^{+}$(b) lies in a slight motion of O$_3$ along $a$-axis in $X_2^{+}$(a) or along $b$-axis for $X_2^{+}$(b).

The fourth unstable mode $X_3^{+}$ ($\omega = 104i$ cm$^{-1}$), degenerated into $X_3^{+}$(a) and $X_3^{+}$(b), consists essentially in a tilting of oxygen octahedra around $a$-axis for $X_3^{+}$(a) and around $b$-axis for $X_3^{+}$(b). In this mode there is no displacement for W atoms while O$_1$ atoms move along the $c$-axis. O$_2$ atoms move along the $b$-axis for $X_3^{+}$(a) or along $a$-axis for $X_3^{+}$(b). 
In the Bi$_2$O$_2$ layer only O$_3$ atoms are moving along $c$-axis.

The fifth unstable mode $X_4^{+}$ ($\omega = 98i$ cm$^{-1}$), degenerated into $X_4^{+}$(a) and $X_4^{+}$(b) is very similar to $X_3^{+}$ except for the motionless O$_3$ atoms  and the displacement of Bi atoms in the Bi$_2$O$_2$ layer.

\subsection{Intermediate and low temperature phases}

The condensation of these different unstable modes within the paraelectric $I4/mmm$ phase will lower the symmetry and decrease the energy. We have seen previously that the internal energy of the $I4/mmm$ phase is significantly higher than that of the $A2/m$ phase but its phonon instabilities are also about twice larger. This let us anticipate that the condensation of the latter could produce a decrease of energy one order of magnitude larger than those reported in the $A2/m$ phase \cite{33} and so yield phases with an internal energy eventually lower that that of the $P\overline{1}$ phase discussed in the previous section. The symmetry resulting from the condensation of individual unstable modes in the $I4/mmm$ phase, together with the decrease in energy with respect to $I4/mmm$ after full atomic relaxation within this given symmetry, are reported in the top part of Table VIII. \textcolor{black} {Although the different unstable modes are twofold degenerated, we restricted ourselves to the condensation of one of the two degenerated modes since other combinations do not appear in the low-symmetry phase experimentally observed that are discussed later.}

The condensation of the polar  $\Gamma_5^{-}$(a) mode in the tetragonal paraelectric $I4/mmm$ phase distorts the structure into a ferroelectric $Fmm2$ (No. 42) phase leading, after full structural relaxation, to cell parameters  a=$5.37\mathring{A}$, b=$5.36\mathring{A}$ and c=$16.00\mathring{A}$.  

The condensation of the $X_2^{-}$ (a) mode leads to $Amma$ (No. 63) symmetry (conventional $CmCm$) and relaxed cell parameters a=$5.36\mathring{A}$, b=$5.34\mathring{A}$ and c=$16.00\mathring{A}$.

 The condensation of $X_2^{+}$ (a) mode leads to a orthorhombic structure $Abam$ (No. 64) (conventional Cmca) with quasi-tetragonal cell parameters a $\approx$ b = $5.23\mathring{A}$ and c = $16.36\mathring{A}$. 

The condensation of $X_3^{+}$ (a) mode leads to an orthorhombic structure  $Bmab$ (No. 64) and relaxed cell parameters a=$5.28\mathring{A}$, b=$5.32\mathring{A}$ and c=$16.26\mathring{A}$. 

Finally, the condensation of $X_4^{+}$(a) leads to $Amaa$ (No. 66) symmetry (conventional Cccm) and relaxed cell parameters   a=$5.28\mathring{A}$, b=$5.31\mathring{A}$ and c=$16.24\mathring{A}$.

 We observed that, amongst these phases, the largest decrease of energy arises from the condensation of the largest $\Gamma_5^{-}$ instability. The related $Fmm2$ phase does however still exhibit phonon instabilities.

The ferroelectric intermediate $B2cb$ phase, observed experimentally, results from the combined condensation of two unstable modes $\Gamma_5^{-}$(a) and $X_3^{+}$(a) in the paraelectric $I4/mmm$ phase, while the addition of a third mode $X_2^{+}$(b) allows the ferroelectric $P2_{1}ab$ ground state to be reached. We notice that a proper (a) or (b) orientation of the unstable modes is necessary to match the experimental structure \cite{16}, with a conventional $a$-axis direction for the polarization. 

The relaxed structural parameters of $B2cb$ and $P2_{1}ab$ phases are presented in Table V and show the same good agreement with experimental data as the other phases previously discussed. \textcolor{black}{For the $B2cb$ phase we notice that, while LDA underestimates the lattice constants, in this case, the GGA calculation of Mohn and Stolen \cite{17} overestimates them clearly.} The related energy, with respect to the $I4/mmm$ phase, are reported in Table VIII. The lowest  internal energy is associated to the  $P2_{1}ab$ phase, which does moreover not exhibit any further phonon instability, in agreement with the fact that it corresponds to the experimental ground state.  

We notice that there are many more ways to combine unstable phonon modes than those producing the $B2cb$ and $P2_{1}ab$ phases and we have explored few of them. We notice for instance that the hypothetical $A2_{1}ma$ phase (No. 26) resulting from the combined condensation of $\Gamma_5^{-}$(a) and $X_2^{+}$(b) mode is never observed experimentally, although it has a lower internal energy than the observed intermediate $B2cb$ phase. The decrease of energy associated to the two phases $B2cb$ and $A2_{1}ma$  is close to the sum of decreases of energy obtained by each individual instability
 producing them, emphasizing the weak interactions between the rotations and the polar mode. At the opposite, the $P2_{1}/a$ phase (No. 14) arising from the condensation of $X_3^{+}$(a) and $X_2^{+}$(b) is much higher in energy emphasizing the fact that both types of rotations tend to exclude each others but not totally. Their combination with the polar mode will so produces the $P2_{1}ab$ ground state.

At the end of Table VIII, we also check the sensitivity of the energy to the relative orientation of the different unstable modes appearing in the $P2_{1}ab$  ground state. We see that replacing the $X_3^{+}$(a) mode by $X_3^{+}$(b) to produce a $P2_1$ phase (No. 4) has a larger effect on the energy, than replacing the $X_2^{+}$(b) mode by $X_2^{+}$(a) to produce a $Pa$ phase (No. 7).
 We notice that both $X_2^{+}$(a) and $X_2^{+}$(b) modes correspond to oxygen rotations about the c-axis and only differ by slight $O_3$ motions along either a- or b-axis.

\subsection{Coupling of lattice modes and energy landscape}

Recently, the coupling of phonon modes in perovskite compounds has known a significant renewal of interest. For instance, the trilinear coupling between the polarization and two rotational modes in layered perovskite, like artificial (SrTiO$_3$)/(PbTiO$_3$) superlattices \cite{34} or natural Ruddlesden-Popper Ca$_3$Ti$_2$O$_7$ compounds \cite{35}, has been identified as a promising road to achieve new and/or enhanced functional properties. Aurivillius phases like SrBi$_2$Ta$_2$O$_9$ are known to exhibit a similar type of trilinear coupling \cite{3} and it is interesting to question if the same type of phenomena is also inherent to Bi$_2$WO$_6$. In what follows, we further quantify the contributions of individual phonon modes to the different phases previously discussed in order to get a better description of the energy landscape and significant phonon couplings.

The  set of phonon eigendisplacements $\eta_i$ of the the $I4/mmm$ phase (normalized such that $\textless \eta \textbar M \textbar \eta \textgreater = 1$)\cite{27} defines a complete basis for the atomic distortions from $I4/mmm$ to a given phase, $X$, so that we can decompose the atomic distortion vector $\Delta$ from $I4/mmm$ to $X$ as a sum of contributions from the different modes $i$: 
\begin{equation}
\Delta = \sum_i Q_i \eta_i 
\end{equation}
where the mode amplitude $Q_i$ is defined as
\begin{equation}
Q_i = A \alpha_i = \textless \eta_i \textbar M \textbar \Delta \textgreater.
\end{equation}
The mode amplitude can be further decomposed into the distortion amplitude $A = \textless \Delta_i \textbar M \textbar \Delta \textgreater$ and the cosine director $\alpha_i$ with respect to distortion direction $\eta_i$ such that $\sum_i \alpha_i^2=1$. The mode amplitude $Q_i$ quantifies which amplitude of $\eta_i$  has been condensed into $I4/mmm$ to reach $X$, while $\alpha_i$ better describes the relative contribution of the different modes entering into a given distortion.

In the top part of Table IX, we report the most significant mode contributions to the atomic distortions in the phases arising from the condensation of individual instabilities (limiting ourselves to contributions $\alpha_i \textgreater 0.01$). We can see that the $Abam$ and $Amaa$ phases arise from the condensation of the related unstable mode only, while $Fmm2$ , $Amma$  and $Bmab$ phases also involve small contributions of additional stable modes of the same symmetry, as induced by phonon-phonon anharmonic couplings.

\begin{table*}
\caption{Contributions $\alpha_i$ (see text) of the different phonon modes $i$ to the atomic distortion from the paraelectric $I4/mmm$ phase to the different intermediate and ground state ferroelectric phases of Bi$_2$WO$_6$. The distortion amplitude A (Bohr) is also reported in each case. Each phonon mode is identified by its symmetry label and frequency (in brackets, in cm$^{-1}$). The corresponding real-space eigendisplacements are reported in Table VII. We limited ourselves to contributions $\alpha_i \textgreater 0.01$.}
\begin{tabular}{lccccccccccccccccccccccccccccccc}
\hline\hline
Phase&&A&&&&$\Gamma_5^{-}$&&$X_2^{-}$&&$X_2^{+}$&&$X_3^{+}$&&$X_4^{+}$&&$\Gamma_5^{-}$&&$X_2^{-}$&& $X_3^{+}$&&$\Gamma_5^{-}$&&&&&&&&  \\
&&&&&&[{\scriptsize198$i$}]&&[{\scriptsize183$i$}]&&[{\scriptsize135$i$}]&&[{\scriptsize104$i$}]&&[{\scriptsize98$i$}]&&[{\scriptsize29}]&&[{\scriptsize80}]&&[{\scriptsize107}]&&[{\scriptsize137}]\\
\hline
Phases arising from single mode condensation  \\
$Fmm2 $&&361.0&&&&0.84&&&&&&&&&&0.51&&&&&&0.11&&\\
$Amma $&&395.6&&&&&&0.79&&&&&&&&&&0.60&&&&&&\\
$Abam $&&444.7&&&&&&&&0.99&&&&&&&&&&&&&&\\
$Bmab $&&444.5&&&&&&&&&&0.98&&&&&&&&0.14&&&&\\
$Amaa $&&384.0&&&&&&&&&&&&0.99&&&&&&&&&&\\
Experimentally observed phases  \\
$B2cb $&&601.0&&&&0.53&&&&&&0.79&&&&0.24&&&&&&0.09&&\\
$P2_{1}ab $&&582.7&&&&0.52&&&&0.59&&0.55&&&&0.04&&&&0.1&&0.09&&\\
Hypothetical phases  & & & & &&&&&&\\
$A2_{1}ma$&&514.5&&&&0.58&&&&0.78&&&&&&0.01&&&&&&0.11&&\\
$P2_1/a$&&455.0&&&&&&&&0.76&&0.62&&&&&&&&0.11&&&&&&&&&&\\ 
\hline\hline
\end{tabular}
\end{table*}

As highlighted in the previous section, the ferroelectric $P2_{1}ab$ ground state results from the condensation of three distinct atomic distortions,  $\Gamma_5^{-}$, $X_3^{+}$ and $X_2^{+}$ into $I4/mmm$.  Due to symmetry considerations, these three modes cannot couple linearly.
However, the Bi$_2$WO$_6$ structure does allow numerous trilinear coupling  terms. Thanks to ISOTROPY software \cite{32}, we have identified the following symmetry-allowed possibilities involving the $\Gamma_5^{-}$ mode and one of the $X_2^{+}$ and $X_3^{+}$ modes :   
\begin{equation*}
\begin{split}
&\Gamma_5^{-} \oplus X_2^{+} \oplus X_3^{-}\\
   &\Gamma_5^{-} \oplus X_2^{+} \oplus X_4^{-}  \\
  &\Gamma_5^{-} \oplus X_3^{+} \oplus X_1^{-} \\
 &\Gamma_5^{-} \oplus X_3^{+} \oplus X_2^{-}\\
\end{split}
\end{equation*}
This teaches us that the presence of $\Gamma_5^{-}$ and $X_3^{+}$ major modes in the intermediate $B2cb$ phase, automatically allow the appearance of minor $X_1^{-}$ and $X_2^{-}$ modes. Moreover the addition of the $X_2^{+}$ mode in the $P2_1ab$ ground state is compatible with the additional appearance of minor $X_3^{-}$ and $X_4^{-}$ modes. If we look in Table X however, to the modes that have been condensed to reach the $B2cb$ and $P2_1ab$ phases, we only see significant contributions from $\Gamma_5^{-}$, $X_3^{+}$ and $X_2^{+}$ modes (the unstable ones and, eventually, higher-frequency modes of the same symmetries). This means that, although various trilinear coupling terms exist, none of them is strong enough to succeed condensing significantly any of the minor modes (i.e. $\alpha_i$ is not rigorously zero but smaller than 0.01). 
   
Since the contributions of all minor modes are negligible, we performed an expansion of the energy (in meV/formula unit) limiting ourselves to the major unstable modes that enter the observed orthorhombic phases only (i.e. $\Gamma_5^{-}$(a), $X_3^{+}$(a) and $X_2^{+}$(b)) :
\begin{equation*}
\begin{split}
&\Delta E_{t} (Q_{\Gamma_5^{-}},Q_{X_2^{+}},Q_{X_3^{+}}) = \\
   &  -2.50 Q_{\Gamma_5^{-}}^{2} - 7.50\times10^{-1} Q_{X_3^{+}}^{2}\\
  &- 1.03 Q_{X_2^{+}}^{2} + 1.36\times10^{-5} Q_{\Gamma_5^{-}}^{4} \\
 & + 1.97\times10^{-6} Q_{X_3^{+}}^{4} +  2.65\times10^{-6} Q_{X_2^{+}}^{4}\\
& + 2.10\times10^{-6} Q_{\Gamma_5^{-}}^{2}Q_{X_3^{+}}^{2} \\
& - 1.79\times10^{-9} Q_{\Gamma_5^{-}}^{2}Q_{X_2^{+}}^{2}\\
& + 1.79\times10^{-6} Q_{X_3^{+}}^{2}Q_{X_2^{+}}^{2} 
\end{split}
\end{equation*}
$Q$ represents the mode amplitude as explained in the previous section . The values of the polynomial coefficients have been obtained by systematic energy calculations after the condensation of various combination of the unstable modes.  

The coefficients of $Q_{\Gamma_5^{-}}^{2}$, $Q_{X_3^{+}}^{2}$ and $Q_{X_2^{+}}^{2}$ are negative, coherently with the phonon instabilities.  The biquadratic coupling coefficient between $\Gamma_5^{-}$(a) and $X_2^{+}$(b) is negative but its value is extremely small.  The biquadratic coupling coefficients between $X_3^{+}$(a) and both $\Gamma_5^{-}$(a) and $X_2^{+}$(b) are positive but still rather small, meaning that although the appearance of $X_3^{+}$(a) tends to exclude the two other mode, this effect is rather small. This is confirmed by the fact that, although $\alpha_{\Gamma_5^{-}}$ decreases due to the appearance of the other modes,  the mode amplitude $Q_{\Gamma_5^{-}}$ remains very similar in $Fmm2$, $B2cb$ and $P2_1ab$ phases. 


The coupling scenario in Bi$_2$WO$_6$ is therefore rather different from those of other previously studied Aurivillius phases, namely, SrBi$_2$Ta$_2$O$_9$ ($m$=2)  and Bi$_4$Ti$_3$O$_{12}$ ($m$=3). In SrBi$_2$Ta$_2$O$_9$, the strongest instability is the octahedral tilting mode $X_3^{-}$ around $a$/$b$ direction and the condensation of this mode alone  results in a nonpolar intermediate phase. This tilting mode and the polar mode $\Gamma_5^{-}$ are mutually exclusive, and the final ferroelectric phase $Ama2_1$ needs the presence of a third stable mode $X_2^{+}$ to stabilize  \cite{3}.  In Bi$_4$Ti$_3$O$_{12}$, as for Bi$_2$WO$_6$, there is three major unstable modes $\Gamma_5^{-}$, $X_2^{+}$ and $X_3^{+}$. However, there is no occurrence of experimentally detected intermediate phases, and the ferroelectric $B1a1$ monoclinic ground state is reached by the simultaneous condensation of the three unstable modes and could be denoted as an avalanche phase transition \cite{4}.

\begin{figure*}[ht]
\centering
\centering\includegraphics[width=17cm , keepaspectratio=true]{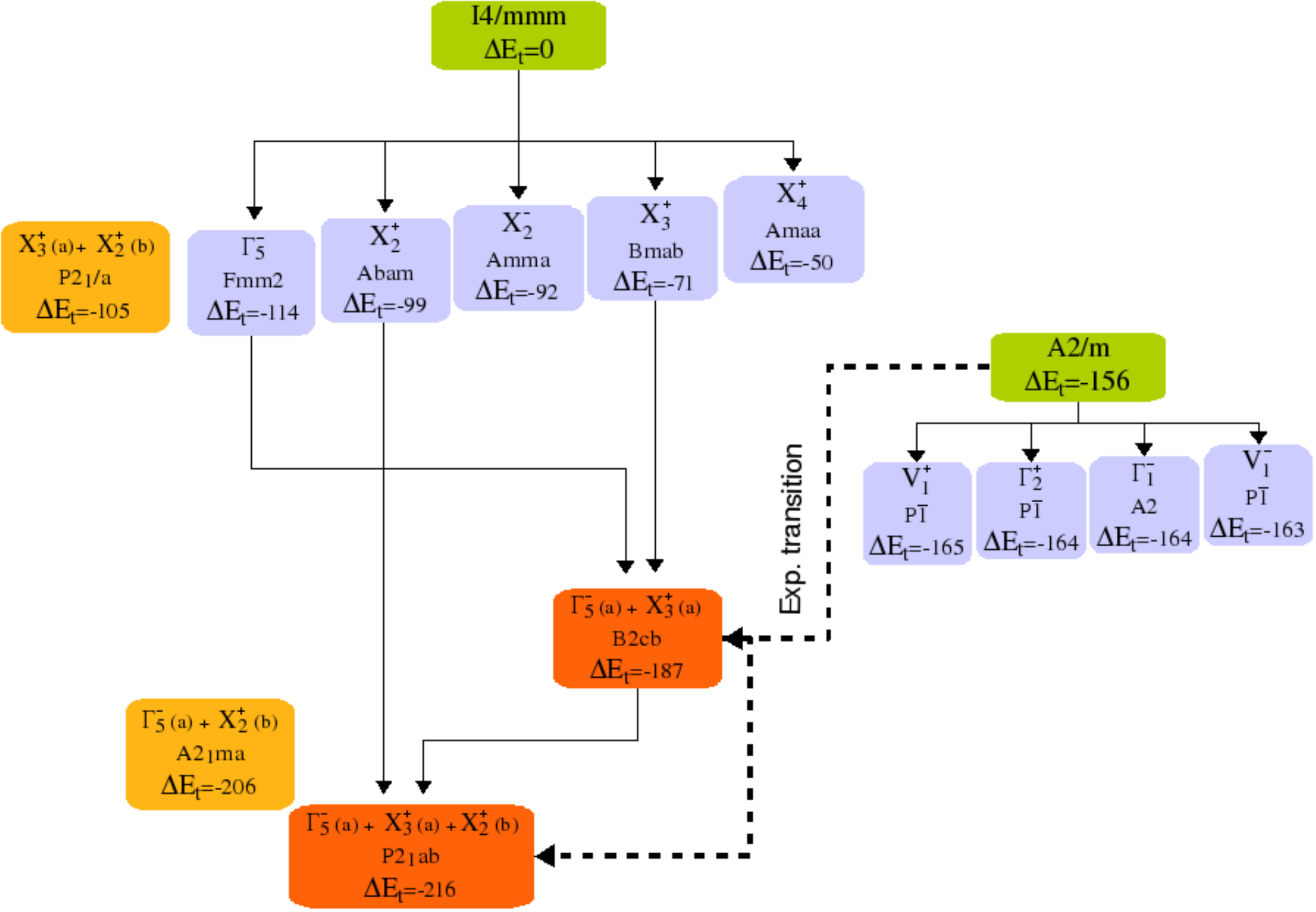}
 \caption{Summary Diagram of phase transitions in Bi$_2$WO$_6$. $\Delta E_{t}$ is in meV/formula unit and represents the difference of energy with respect to $I4/mmm$ taken as the zero energy reference.}
\end{figure*}
 
\subsection{Spontaneous polarization}
\begin{table}
\caption{Spontaneous polarization (in $\mu$C/cm$^2$) of different observed or hypothetical polar phases of Bi$_2$WO$_6$. $P_s(BP)$ refers to the polarization computed using the Berry Phase approach (Eq.1). $P_s(Z^*)$ corresponds to the polarization deduced from the knowledge of the Born effective charges and atomic displacements (Eq.2). The latter has been decomposed into the contributions coming from the lowest frequency $\Gamma_5^-$ polar modes \cite{36}, identified by their frequency $\Omega$ (in brackets, in cm$^{-1}$) and mode polarity $\frac{e}{\Omega}$ $\sum$ ${Z}^*$ $\eta$ (in parenthesis, in $\mu$C/cm$^2$). }
\begin{tabular}{cccccccccccccccccccccccccccccccc}
\hline\hline
Phase&&&&$\Gamma_5^{-}$&&$\Gamma_5^{-}$&&$\Gamma_5^{-}$&&&&$P_s(Z^*)$&&$P_s(BP)$  \\
&&&&[{\scriptsize198$i$}]&&[{\scriptsize29}]&&[{\scriptsize137}]&&\\
&&&&(0.18)&&(0.03)&&(-0.06)&&\\
\hline
$Fmm2 $&&&&55.26&&4.82&&-2.40&&&&58.59&&53.98\\
$B2cb $&&&&57.92&&3.77&&-3.30&&&&58.29&&52.06\\
$P2_{1}ab $&&&&55.09&&0.61&&-3.17&&&&52.63&&48.16\\
$A2_{1}ma$&&&&53.32&&-0.13&&-3.42&&&&49.96&&46.98\\
\hline\hline
\end{tabular}
\end{table}

In both orthorhombic ferroelectric phases, the spontaneous polarization $P_s$ expands along $a$-direction only, with no component along $c$. We report, in Table X, the spontaneous polarizations of different observed and hypothetical phases as obtained using the Berry phase approach ($P_s(BP)$) and using Eq.(2) and the knowledge of the Born effective charges and atomic distortions ($P_s(Z^*)$). Again, determination of $P_s(BP)$ requires careful determination of the polarization quantum (see discussion in Section III-D). 

In Table X we also report the contributions to the full polarization arising from the individual ${\Gamma_5^-}$ polar modes. We see that the main contribution arises from the unstable ${\Gamma_5^-}$ mode, which is also the most polar.  We observe that the spontaneous polarization is rather constant in the different phases, in agreement with the weak coupling between polar and rotational modes, yielding an almost constant amplitude of $Q_{\Gamma_5^-}$ as discussed in the previous Section. The polar mode of highest frequency has a tiny contribution but tends to reduce the full polarization. We notice that $P_s(Z^*)$ slightly overestimates $P_s(BP)$ , but since the same distortion is used in both approaches this has to be linked to the evolution of the Born effective charges along the path of atomic distortion and neglected in the computation of $P_s(Z^*)$.

Independently, we also decomposed the polarization in terms of contributions coming from the individual layers: the main contribution (76 $\%$  of total polarization) comes from the perovskite-like block, due to the displacement of W opposite to its surrounding octahedral oxygen atoms  O$_1$ and O$_2$ while a smaller part (24 $\%$ of total polarization) comes from the Bi$_2$O$_2$ layer, due to Bi displacement opposite to O$_3$. This means that the contribution of Bi$_2$O$_2$ to the polarization is not dominant, contrary to the usual belief \cite{6}.

\section{Discussion}

In the previous Sections, we have discussed many different phases of Bi$_2$WO$_6$. In Figure 4, we provide a summary comparing the internal energies of all the phases we have studied. This diagram gives a global view and helps clarifying the phase transitions.

At high temperature Bi$_2$WO$_6$ crystallizes in a monoclinic $A2/m$ structure. Starting from this configuration, displacive phase transitions are possible to $A$2 or $P\overline{1}$ phases. However, Bi$_2$WO$_6$ can adopt another polymorphic $I4/mmm$ tetragonal form, which displays much stronger phonon instabilities, yielding orthorhombic $B2cb$ and $P2_1ab$ phases of lower energies. Starting at high temperature from the $A2/m$ structure, the system has to exhibit a reconstructive phase transition to reach the $P2_1ab$ ground-state. 

\section{Conclusions}

We have performed a first-principles study of Bi$_2$WO$_6$ using Density Functional Theory  within the Local Density Approximation. 

First, we characterized the high-temperature paraelectric monoclinic $A2/m$ phase. Phonon frequencies and dielectric properties were computed. This monoclinic structure,  with a configuration mixing edge-sharing and corner-sharing (WO$_4$)$^{-2}$ octahedra, exhibits like in typical perovskites anomalous Z* on W and its surrounding oxygens along the corner-sharing chains, yielding potentially a non-negligible spontaneous polarization along this direction.  Lattice dynamics study reveals four structural insabilities that lead the system to energetically quasi-equivalent monoclinic phases. One of these phases is ferroelectric with space group $A$2 and a spontaneous polarization of 20 $\mu$C/cm$^2$. 

Second, we considered a hypothetical paraelectric $I4/mmm$ phase that, although not observed experimentally, has a symmetry that is the supergroup of the low temperature orthorhombic phases, observed experimentally. We fully characterized the $I4/mmm$ phase, that exhibits five degenerated phonon instabilities.  Our calculations agree with the observation  that Bi$_2$WO$_6$ has a $P2_1ab$ ground-state structure. This ground state can be reached from the  condensation of three unstable phonons in the  paraelectric $I4/mmm$ phase : a polar mode, $\Gamma_5^{-}$, a rotation of oxygen octahedra around $c$-axis, $X_2^{+}$ and a tilt of oxygen octahedra around $a$-axis, $X_3^{+}$. The experimentally observed $B2cb$ phase appears as an intermediate phase along this path, arising from the condensation of the $\Gamma_5^{-}$ and $X_3^{+}$ modes. Our calculations reveal a weak coupling between the different modes entering the ground-state and the absence of significant distortion arising from minor modes allowed by symmetry. 
As in other perovskite ferroelectric oxides, the Z* are anomalously large, yielding a strong spontaneous polarization of 48 $\mu$C/cm$^2$ in the $P2_1ab$ phase. The significantly lower internal energy of the $P2_1ab$ phase, compared to the $P\overline{1}$ phase, is in agreement with the fact that starting from $A2/m$ phase, the system displays a reconstructive phase transition.

\section{Acknowledgment}
This work was performed during a visit of H.D. at the University of Li\`ege in Belgium with the combined support of the Algerian Ministry of High Education and Scientific Research (MESRS) and the Interuniversity Attractive Pole Program from the Federal Science Research Policy of Belgium. Ph.G. thanks the Francqui Foundation for Research Professorship. E.B. thanks the FRS-FNRS Belgium.

\section{Appendix}
           
In this Section we report for completeness the full set of $\Gamma$ phonons in the $A2/m$ (see Table XI) and $I4/mmm$ phases (see Table XII) of Bi$_2$WO$_6$ as computed within the LDA in the fully relaxed structures. For the $A2/m$ phases no previous calculations have been reported and no experimental data are available. Our results therefore constitute a theoretical prediction that might be useful to interpret future experimental measurements. For the hypothetical $I4/mmm$ phase, our results are compared to those of Machado {\it et al.} \cite{16} also within the LDA but at a larger volume. As usual in ferroelectric oxides, the use of a larger volume favors the ferroelectric instability.  

\begin{table*}
\caption{Frequencies (in cm$^{-1}$) of the phonon modes at $\Gamma$ in the monoclinic $A2/m$ phase of Bi$_2$WO$_6$.}
\begin{tabular}{lccccccccccccccccccccccc}
\hline\hline
& & &$\Gamma_1^{-}$ ($A_u$)& &$\Gamma_1^{+}$ ($A_g$)& & $\Gamma_2^{-}$ ($B_{u}$)& &$\Gamma_2^{+}$ ($B_{g}$)  \\
\hline
&&&108$i$ &&87&&54 &&63$i$   \\
&&&61 &&91 &&82&&178   \\
&&&278&& 122&&96&& 241       \\
&&&295 &&160&&144 &&303     \\
&&&336&&167&&194&&378\\
&&&408&&192&&209&&455\\
&&&653&&272&&280&&456\\
&&&654&&298&&309&&\\
&&&&&333&&352&&\\
&&&&&380&&353&&\\
&&&&&404&&358&&\\
&&&&&441&&444&&\\
&&&&&468&&473\\
&&&&&497&&624\\
&&&&&642&&787\\
&&&&&663&&876\\
&&&&&665\\
&&&&&750\\
&&&&&895\\
\hline\hline
\end{tabular}
\end{table*}

\begin{table*}
\caption{Frequencies (in cm$^{-1}$) of the phonon modes at $\Gamma$ in the tetragonal $I4/mmm$ phase of Bi$_2$WO$_6$. Values between brackets are those previously reported in Ref. \cite{16}.}
\begin{tabular}{cccccccccccccccccccc}
\hline\hline
& & &$\Gamma_5^{-}$ ($E_u$)&&$\Gamma_5^{-}$ ($E_u$)[16]&&$\Gamma_5^{+}$ ($E_g$)& & $\Gamma_3^{-}$ ($A_{2u}$)&&$\Gamma_3^{-}$ ($A_{2u}$)[16]& &$\Gamma_1^{+}$ ($A_{1g}$)&&$\Gamma_4^{-}$ ($B_{2u}$)&& $\Gamma_2^{+}$ ($B_{1g}$) \\
\hline
&&&198$i$&&(271$i$)&&71&&89&&(51)&&176&&312&&437   \\
&&&29&&(34$i$)&&281&&348&&(337)&&767&&&&           \\
&&&137&&(133)&&517&&469&&(437)&&&&&&                   \\
&&&359 &&(211)&&&&674&&(630)&&&&&&              \\
&&&650 &&(525)&&&&&&&&&            \\

\hline\hline
\end{tabular}
\end{table*}


\begin{thebibliography}{40}
\bibitem{note}
Real-space eigendisplacements $\eta$ are related to the dynamical matrix eigenvectors $\gamma$ through $\eta_{\kappa,\alpha} = \gamma_{\kappa,\alpha} / \sqrt(M_{\kappa})$ where $\kappa$ runs over all the atoms, $\alpha$ runs over the three directions and $M_{\kappa}$ is the mass of atom ${\kappa}$. Since dynamical matrix eigenvectors are normalized such as $\textless \gamma \textbar \gamma \textgreater = \sum_{\kappa,\alpha} \gamma_{\kappa,\alpha}^*\gamma_{\kappa,\alpha}=1$, the related eigendisplacements are normalized such as $\textless \eta \textbar M \textbar \eta \textgreater =\sum_{\kappa,\alpha} M_{\kappa}\eta_{\kappa,\alpha}^*\eta_{\kappa,\alpha} =1$ where $M$ is the mass matrix. \cite{27,PhD}
\bibitem{33}
Assuming a double-well internal energy, related to the ionic degrees of freedom $Q$ and the unstable phonon frequency $\omega i$, of the form : $E= (-1/2) \omega^2  Q^2 + (1/4) \beta Q^4$, the well depth is $\Delta E=\omega^4/4\beta$. So, increasing the amplitude of the instability $\omega$ by a factor of 2, will increase the well depth by a factor of 16 (assuming the anharmonic coefficient unchanged).
\bibitem{36}
Contributions from other polar modes are not reported.
\end{thebibliography}
\end{document}